\documentclass[twocolumn]{jpsj2} 
\title{Transition Phenomena Induced by Internal Noise and Quasi-absorbing State}

\author{Jun \textsc{Ohkubo}$^{1}$\thanks{E-mail address: ohkubo@issp.u-tokyo.ac.jp},
Nadav \textsc{Shnerb}$^{2}$\thanks{E-mail address: shnerbn@mail.biu.ac.il}
and David A. \textsc{Kessler}$^{2}$\thanks{E-mail address: kessler@dave.ph.biu.ac.il}
}

\inst{
$^{1}$Institute for Solid State Physics, University of Tokyo, 
Kashiwanoha 5-1-5, Kashiwa, Chiba 277-8581\\
$^{2}$Department of Physics, Bar-Ilan University,
Ramat-Gan 52900, Israel
}

\abst{
We study a simple chemical reaction system and effects of the internal noise.
The chemical reaction system causes the same transition phenomenon discussed
by Togashi and Kaneko [Phys. Rev. Lett. {\bf 86} (2001) 2459; J. Phys. Soc. Jpn. {\bf 72} (2003) 62].
By using the simpler model than Togashi-Kaneko's one,
we discuss the transition phenomenon by means of a random walk model and an effective model.
The discussion makes it clear that 
quasi-absorbing states, which are produced by the change of the strength of the internal noise,
play an important role in the transition phenomenon.
Stabilizing the quasi-absorbing states causes
bifurcation of the peaks in the stationary probability distribution discontinuously.
}

\kword{stochastic process, transition phenomena, biochemical systems, Random walk model, Fokker-Planck equation}

\begin{document}
\maketitle

\section{Introduction}

Noise has many roles in many stochastic processes,
and the analysis of the behavior of complex stochastic systems
is one of the most interesting issues in statistical physics and related areas.
Chemical reactions are described traditionally in terms of kinetic rates,
and a deterministic rate equation approach is often used.
While effects of noise or fluctuation can not be treated directly by the deterministic rate equation approach,
a stochastic system could cause a drastic modification of macroscopic properties due to the 
noise and fluctuation effects.

For example, there is a simple model for the transition phenomenon induced by external noise
\cite{Horsthemke1984}.
The behavior of the following chemical reaction system
\begin{align}
\begin{array}{l}
A + X + Y \leftrightarrows 2Y + A^* \\
B + X + Y \leftrightarrows 2X + B^*
\end{array}
\label{reaction_external}
\end{align}
can be represented adequately by a deterministic phenomenological equation
\begin{align}
\frac{\mathrm{d}X}{\mathrm{d}t} = \alpha - X + \lambda X (1-X),
\label{eq_external}
\end{align}
where the variable $X$ is the concentration of chemical substance $X$ in the chemical reaction system
\eqref{reaction_external}.
When $A^*$ and $B^*$ are in large excess so that their fluctuations can be neglected,
$\alpha$ becomes a constant.
The fluctuation effects of $A$ and $B$ are included in $\lambda$.
If we neglect the fluctuations of $A$ and $B$, the coefficient $\lambda$ seems to be a constant, 
and then eq.~\eqref{eq_external} 
describes a deterministic motion of a chemical substance $X$.
In this case, eq.~\eqref{eq_external} gives only one stable solution.

In order to include the fluctuation effect of $A$ and $B$,
one assumes that the external fluctuations are extremely rapid and then 
the effects are written by $\lambda = \lambda^* + \sigma \xi$,
where $\xi$ means a white Gaussian noise with zero mean and variance $1$.
For the sake of simplicity, we take $\alpha = 1/2$ and $\lambda^* = 0$.
In this case, the stationary probability density has one peak or two peaks
depending on the value of $\sigma$ \cite{Horsthemke1984};
\begin{align}
\begin{cases}
X^\textrm{peak} = \frac{1}{2} & \textrm{for} \,\, \sigma^2 < 4, \\
X_{\pm}^\textrm{peak} = \frac{1}{2} \left[ 1 \pm (1- 4 / \sigma^2 )^{1/2} \right]
& \textrm{for} \,\, \sigma^2 > 4 ,
\end{cases}
\end{align}
i.e., when the fluctuation is small ($\sigma^2 < 4$),
the probability density has only one peak,
and in contrast the probability density has two peaks if the fluctuation is large ($\sigma^2 > 4$).
The change of the number of peaks means the transition phenomenon induced by the external noise \cite{Horsthemke1984}.

As the above example shows, 
a deterministic equation is not enough to treat a stochastic process adequately.
There are many examples for such stochastic processes,
e.g., a chemical reaction system, a prey-predator system, disease spreading system, an so on.
These systems are described by a reaction scheme such as eq.~\eqref{reaction_external},
and then they consist of discrete components such as molecules (chemical substances) and individuals.
So far, the importance of the discreteness has been pointed out by several authors
\cite{Rao2002,Brunet1997,Kessler1998,Shnerb2000,Shnerb2001,Bettelheim2001,Marion2002,Abta2007,Tauber2005}.

Togashi and Kaneko \cite{Togashi2001,Togashi2003,Togashi2007}
have shown that a novel transition phenomenon occurs in a chemical reaction system with small size.
In their work, a small autocatalytic system, which consists of at least four chemical substances,
has been investigated.
Togashi and Kaneko have shown that there is a nontrivial transition phenomenon induced by the molecular discreteness.
However, the models in references 11, 12, and 13 are
a little complicated to be treated analytically,
so that only numerical experiments have been performed in order to research the transition phenomenon.
In order to investigate the transition phenomenon quantitatively 
and to make an intuitive picture for the transition phenomenon,
it would be needed to construct a simple model which is easier to treat analytically. 

In the present paper, 
we propose a simpler model than Togashi-Kaneko's one; the simple model shows the same transition phenomenon
which was discussed by Togashi and Kaneko.
The model consists of only two chemical substances, and hence 
we can treat the model analytically with a certain assumption.
By numerical experiments, we confirm that the same transition phenomenon occurs even in our simple model.
Furthermore, using an assumption that there is a restriction of the total number of molecules,
it is possible to write down the master equation with only one variable.
The master equation is easy to be analyzed,
and the transition point can be estimated adequately.
In addition, we can construct a random walk model from the master equation,
and the random walk model gives us intuitive pictures of the role of the internal noise.
By the intuitive pictures,
we conclude that quasi-absorbing states produced by the internal noise play an important role
for the transition phenomenon, rather than the discreteness effects.
In order to check this fact, we use an effective chemical reaction model
and analyze it by the Fokker-Planck equation approach;
the effective chemical reaction model can cause the same transition phenomenon
even in the large system size.

The present paper is organized as follows.
In \S~2, we propose an autocatalytic model which shows a transition phenomenon induced by the internal noise,
and give results of numerical experiments.
Section~3 gives an analytical treatment for the autocatalytic model.
In addition, we propose an effective chemical reaction model,
which clarifies the quasi-absorbing states are important rather than the discreteness.
Finally, we give some concluding remarks in \S~4.

\section{Autocatalytic model}

\subsection{Model}

\begin{figure}
\begin{center}
  \includegraphics[width=85mm,keepaspectratio,clip]{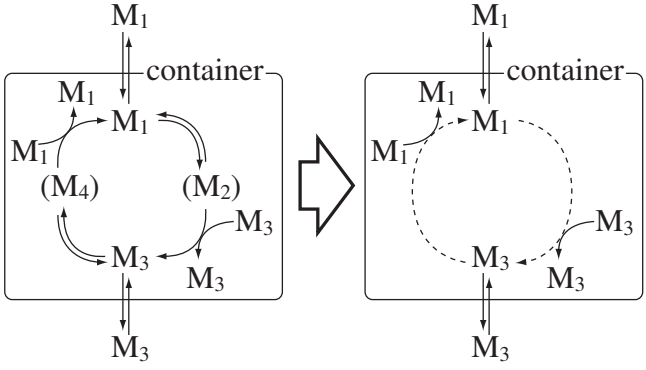} 
\caption{
Illustrative explanation of an autocatalytic process in a container.
A container is in contact with a reservoir of molecules, and the chemicals $M_1$ and $M_3$ 
diffuse in and out through the membrane of the container.
The reaction network in the left side can be reduced to that in the right side
with some assumptions.
}
\label{fig_reaction}
\end{center}
\end{figure}

We consider a simple chemical reaction network shown in Fig.~\ref{fig_reaction}.
We denote the number of chemical substance $M_i$ as $N_i$,
and the concentration of each chemical substance as $x_i \equiv N_i / V$,
where $V$ is the volume of the container.
Although there are four chemical substances $M_1$, $M_2$, $M_3$, $M_4$ in the chemical reaction network 
in Fig.~\ref{fig_reaction},
the chemical reaction network would be described adequately only with two variables $N_1$ and $N_3$
by introducing some assumptions.

For example, we consider the following additional reaction for the reaction $M_1 \leftrightarrows M_2$:
$M_1 + M_\textrm{TMP} \leftrightarrows M_2$.
We assume that the chemical substance $M_\textrm{TMP}$ is in large excess,
and the reaction $M_1 + M_\textrm{TMP} \leftrightarrows M_2$ occurs very rapidly.
Because of the rapid attachment and detachment of $M_\textrm{TMP}$,
we consider that the number of chemical substance $M_2$ is the same as that of $M_1$, i.e., $N_1 = N_2$.
When the molecule $M_3$ is near the molecule $M_1$,
the chemical substance $M_\textrm{TMP}$ would attach to the molecule $M_1$ rapidly, and
the reaction with $M_2 + M_3$ occurs.
As for the chemical substance $M_4$, we also assume the similar assumption, i.e., $N_3 = N_4$.
As a result, we presume the reaction network in the left side of Fig.~\ref{fig_reaction}
can be reduced to that in the right side of Fig.~\ref{fig_reaction}.

Using the above discussions, the autocatalytic chemical reaction can be described simply as
\begin{align}
\begin{cases}
M_1 + M_3 \to  2 M_3   \\ \qquad \textrm{with rate}\,\, r_1 x_1 x_3 V = r_1 N_1 N_3 / V,\\
M_3 + M_1 \to  2 M_1   \\ \qquad \textrm{with rate}\,\, r_3 x_1 x_3 V = r_3 N_1 N_3 / V,\\
\textrm{(outside)} \to M_i \\ \qquad \textrm{with rate}\,\, D_i V s_i \,\, (i = 1,3),\\
M_i \to \textrm{(outside)} \\ \qquad \textrm{with rate}\,\, D_i V x_i = D_i N_i \,\, (i = 1,3),
\end{cases}
\label{reaction}
\end{align}
where $r_i$ is the reaction rate, $D$ the diffusion rate across the surface of the container,
and $s_i$ the concentration of the molecule in the outside of the container.
A container is in contact with a reservoir of molecules, and the chemical substances $M_1$ and $M_3$ 
diffuse in and out through the membrane of the container \cite{note}.
Although the reaction network might be artificial,
the model is useful in order to investigate the effects of the internal noise.

In what follows, we assume that $r_1 = r_3 = 1$, $s_1 = s_3 = 1$ and $D_1 = D_3 \equiv D$ for simplicity,
while the transition phenomena to be presented in the paper will persist if we drop these conditions.
Because we set $r_1 = r_3$, 
the rate with which the chemical substance $M_3$ is generated by the autocatalytic reaction
is the same as the rate for the generation of the chemical substance $M_1$.
Hence, the deterministic rate equation is denoted by
\begin{align}
\frac{\mathrm{d}x_i}{\mathrm{d}t} = D(1 - x_i), \quad (i = 1, 3).
\label{eq_deterministic}
\end{align}
The deterministic rate equation would be valid when one takes a continuum limit, given by $V \to \infty$.
In this limit, the fluctuation of concentration $x_i$ is negligible.
Obviously, the deterministic rate equation means that the fixed point is $x_i = 1$.

\subsection{Numerical results}

The chemical reaction system \eqref{reaction} consists of a Markov jump process in continuous time.
Such dynamics can be simulated (exactly) on a computer using standard discrete-event simulation techniques.
The most standard implementation of this strategy is known as the Gillespie algorithm \cite{Gillespie1977}.
In the Gillespie algorithm, the lapse time to the next event is determined 
by exponentially distributed random numbers,
and one determines which event occurs depending on the rate of the event.

We apply the Gillespie algorithm to the chemical reaction network \eqref{reaction},
and study the transition phenomenon induced by the internal noise and the discreteness numerically.
In the whole numerical experiments, we set $D = 1 / 64$,
and investigate the effect of the change in the volume $V$.

\begin{figure}[t]
\begin{center}
  \includegraphics[width=90mm,keepaspectratio,clip]{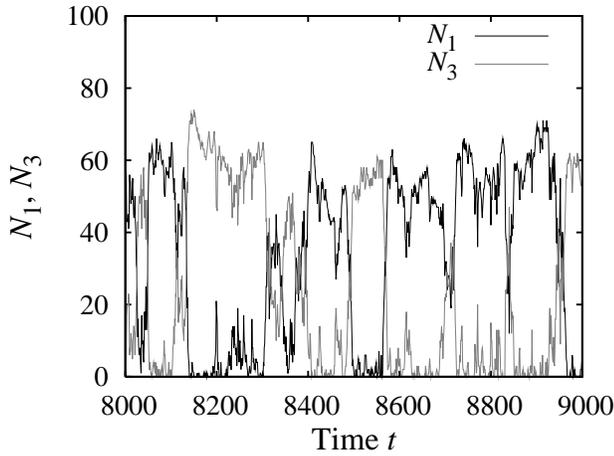} 
\caption{
Sample paths of the number of molecules $N_1$ and $N_3$.
Here $s_1 = s_3 = 1$, $D = 1/64$ and $V = 32$.
The paths fluctuate around $N_i = 0$ or $N_i = 64$.
}
\label{fig_sample_path}
\end{center}
\end{figure}

We here investigate the time evolution of the number of molecules $N_1$ and $N_3$
in the case with $V = 32$.
One might expect that the concentrations $x_1$ and $x_3$ fluctuate around the outside concentration,
which is expected by the deterministic rate equation approach.
However, Fig.~\ref{fig_sample_path} shows a different behavior;
the paths seem to fluctuate around $N_i = 0$ or $N_i = 64$.

\begin{figure}
\begin{center}
  \includegraphics[width=90mm,keepaspectratio,clip]{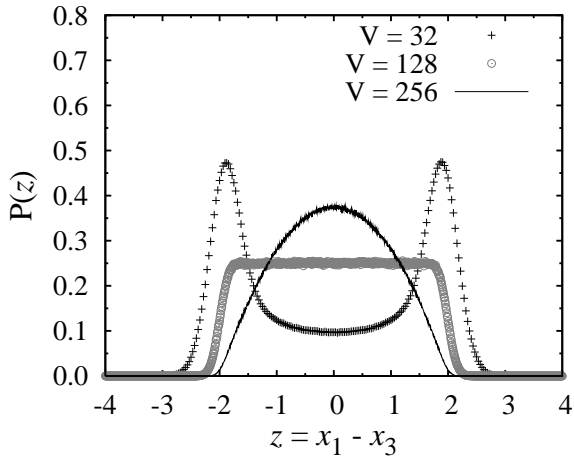} 
\caption{
The probability distribution of $z \equiv x_1 - x_3$.
Here $D = 1/64$.
When the system size is large ($V > 128$), there is only one peak.
In contrast, there is two peaks at $z = \pm 2$ for the small system size ($V < 128$).
}
\label{fig_prob_z}
\end{center}
\end{figure}

In order to clarify the unexpected behavior of $N_1$ and $N_3$ ($x_1$ and $x_3$),
we calculate the probability distribution of the difference between two concentrations, $z \equiv x_1 - x_3$.
As anticipated by the deterministic rate equation~\eqref{eq_deterministic},
when the system volume is large the concentration of chemical substances $M_1$ and $M_3$ is determined
by the outside concentration $s_1 = 1$ and $s_3 = 1$.
In the present case, $z$ should tend to take the value around zero, because $s_1 = s_3$.
Figure~\ref{fig_prob_z} shows the probability distribution of $z$.
When the system volume $V$ is large, the probability distribution has a peak at $z = 0$.
On the other hand, the probability distribution has two peaks if the system volume $V$ is small.
The two peaks around $z = \pm 2$ mean that the number of only one chemical substance $M_1$ or $M_3$
fluctuates around $(s_1+s_3) V$, and that of the other chemical substance is nearly zero
(see Fig.~\ref{fig_sample_path}).
Note that the peak around $z = 0$ becomes smaller and smaller as the volume $V$ decreases,
and the two peaks around $z = \pm 2$ seem to emerge suddenly;
this is different from the transition phenomenon of the reaction network \eqref{reaction_external}
in \S~1, in which one peak gradually splits into two peaks.

As stated above, one of the features of the transition phenomenon
is the discontinuous split of peaks, which is different from the transition phenomenon
induced by the external noise.
The feature has also been observed in Togashi-Kaneko model.
Togashi-Kaneko model in reference 11 has four chemical substances,
so that Togashi-kaneko model might have different characters from our simple model.
However, we consider that the scenario for the transition is the same;
the decrease of the volume causes the increase of the internal noise,
and then the discontinuous split occurs.

\section{Discussions for the transition phenomenon}

\subsection{Master equation approach}

In the previous section, we have numerically confirmed the phenomenon induced by the internal noise.
In this section, we try to treat the model analytically,
because analytic treatments for a simple toy model would give us intuitive pictures for the transition phenomenon.

In order to investigate the chemical reaction network introduced in \S~2 analytically,
we introduce an assumption such that ``the total number of molecules is conserved.''
The assumption means that $N \equiv N_1 + N_3 = (s_1 + s_3)V$ is time-independent.
From the assumption, we suppose that the outflow of one $M_1$ molecule immediately induces
the inflow of one $M_3$ molecule, and so forth.
We therefore derive the following master equation for $M_1$ ($n \equiv N_1$):
\begin{align}
\frac{\mathrm{d} P(n)}{\mathrm{d}t} =&
F(n) + G(n), 
\label{eq_master}
\end{align}
where
\begin{align}
F(n) =& \frac{1}{V}  (n+1) \{ (s_1+s_3) V - (n+1) \} P(n+1) \notag \\
&+  \frac{1}{V} (n-1) \{(s_1+s_3)V - (n-1)\} P(n-1) \notag \\
&- \frac{2}{V} n \{(s_1+s_3) V - n \} P(n),
\end{align}
and
\begin{align}
G(n) =& D' (s_3 V + n + 1) P(n+1) \notag \\
&+ D' \{ (2s_1 + s_3) V - (n-1) \} P(n-1) \notag \\
&- 2 D' (s_1 + s_3) V P(n).
\end{align}
Note that the conservation of the total number of molecules effectively changes
the diffusion constant $D$, and hence we denote it as $D'$.
The reason why we divided the master equation into two parts ($F(n)$ and $G(n)$)
is explained later.

The transition point $V_\textrm{c}$ is estimated as follows.
Here, we define the transition as the emergence of the peak at $n = 0$.
In order to calculate the transition point with the parameters $s_1 = s_3 = 1$,
we use the detailed balance between the states $n = 0$ and $n = 1$:
\begin{align}
\frac{1}{V} (2 V - 1) P(1) +  D' (V+1) P(1) = 3 D' V P(0).
\end{align}
The emergence of the peak at $n = 0$ is characterized by the fact of $P(0) > P(1)$.
Hence, the transition point is determined by the condition $P(0) = P(1)$, and 
we estimate the transition point $V_\textrm{c}$ as $V_\textrm{c} = 1 / D'$.
The conservation of the total number of molecules means that the diffusion of $M_1$ becomes larger; 
the diffusion of $M_3$ causes the diffusion effect for $M_1$ due to the conservation.
The effective diffusion constant should therefore be set as $D' \simeq D / 2$, and we obtain
\begin{align}
V_\textrm{c} \simeq 2 / D.
\end{align}
In our case, $V_\textrm{c} \simeq 128$ because of $D = 1 / 64$.
This result is consistent with the numerical results in Fig.~\ref{fig_prob_z}.

\begin{figure}
\begin{center}
  \includegraphics[width=60mm,keepaspectratio,clip]{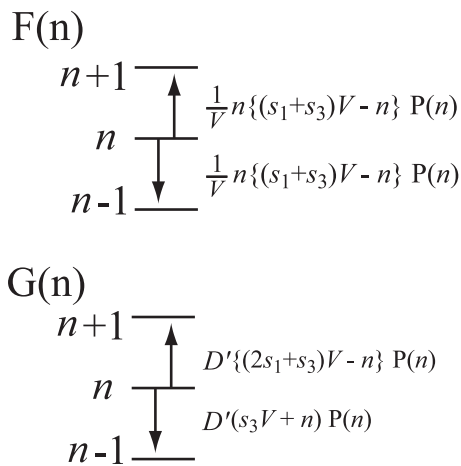} 
\caption{
The difference of the transition rates between $n \to n+1$ and $n \to n-1$.
See the master equation \eqref{eq_master}. 
}
\label{fig_transition}
\end{center}
\end{figure}

Analyzing the master equation \eqref{eq_master},
we obtain a simple random walk picture.
In order to make an intuitive picture, 
we investigate the difference of the transition rates between $n \to n+1$ and $n \to n-1$.
Figure~\ref{fig_transition} shows that the transition rates.
The first term of eq.~\eqref{eq_master}, $F(n)$, corresponds to a random-fluctuating force,
because the transition rate of $n \to n+1$ is the same as that of $n \to n-1$.
On the contrary, the second term of eq.~\eqref{eq_master}, $G(n)$ has an interesting property.
As shown in Fig.~\ref{fig_transition},
there is a difference between the transition rate $n \to n+1$ and that of $n \to n-1$;
we define the difference as $E(n)$ and then
\begin{align}
E(n) = 2D' P(n) (s_1 V - n).
\end{align}
Note that the difference $E(n)$ changes its sign due to the value of $n$;
when $n < s_1 V$, there is a flow from $n-1$ to $n+1$.
The change of the sign of $E(n)$ means that there is an attracting force
toward $n = s_1 V$.
Hence, we will have a simple potential picture for the effect of the term $G(n)$.

We summarize the characteristics of the random walk model:
\begin{enumerate}
\item There is an attracting force which pulls back the number of molecules $n$ into the outside on $s_i V$.
\item There is an internal noise which does not appear in the deterministic rate equation~\eqref{eq_deterministic}.
The internal noise changes the number of molecules $n = N_1$ into $n \pm 1$ with the rate $2 N_1 N_3 / V$.
\end{enumerate}

\begin{figure}
\begin{center}
  \includegraphics[width=70mm,keepaspectratio,clip]{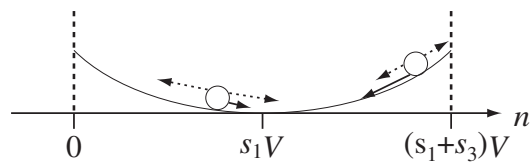} 
\caption{
Random walk model for transition phenomenon induced by the internal noise.
When we assume that the total number of molecule $M_1 + M_3$ is conserved,
the chemical reaction in \S~2 is reduced into the random walk model.
There are two forces acting on the random walker:
one is an attracting force which pulls the random walker toward the potential minimum (solid arrow),
and the other is a fluctuating force (dashed arrow).
}
\label{fig_model}
\end{center}
\end{figure}

In Fig.~\ref{fig_model}, we show the intuitive picture for the random walk model.
The position of the random walker, $n$, represents the number of molecules $N_1$.
The random walk model has two forces:
one is the attracting force which pulls the position of the random walker toward the potential minimum;
the other is a random fluctuation and shows the effects of the internal noise.
The dashed arrow in Fig.~\ref{fig_model} shows the random fluctuation,
and the solid arrow means the attracting force.
Considering the characteristics of the random walk model,
we conclude that the random fluctuation becomes smaller 
and the attracting force becomes larger as $n$ approaches the boundaries ($0$ or $(s_1+s_3)V$).

From the random walk model, 
we have intuitive pictures for the transition phenomenon induced by the internal noise.
The internal noise plays an important role in order to escape from the potential minimum ($n = s_1 V$)
because the attractive force immediately pulls the random walker toward the potential minimum
if the random fluctuating force is small.
However, the decrease of the internal noise near the boundary would be also important.
When the random walker becomes near the boundary, 
the random fluctuating force becomes smaller than that near the potential minimum,
and hence the random walker tends to stay in such regions for a longer time
compared with the region near $n = s_1 V$.

\subsection{Effective model}

Considering the random walk model carefully,
we can construct a most natural one variable model which causes the transition phenomena.
The chemical reaction model is as follows:
\begin{align}
\begin{array}{l}
A  \stackrel{k}{\to} B, \\
B  \stackrel{k}{\to} A, \\
A + B  \stackrel{s}{\to} 2A, \\
A + B  \stackrel{s}{\to} 2B. 
\end{array}
\label{reaction_simple}
\end{align}
The rate equations for this model are
\begin{align}
\begin{array}{l}
\frac{\mathrm{d}}{\mathrm{d}t} A = -k A + k B, \\
\frac{\mathrm{d}}{\mathrm{d}t} B = k A - k B, 
\end{array}
\end{align}
where we denote the number of particles $A$ (or $B$) by the same symbol $A$ (or $B$).
When we assume that the total number of particles, which we denote by $N (\equiv A + B)$, is conserved, 
the rate equations are reduced to the following equation:
\begin{align}
\frac{\mathrm{d}}{\mathrm{d}t} A = k (N - 2 A).
\end{align}
The deterministic dynamics is driven to the equally mixed state.
The $s$ reaction, on the other hand, leads to a segregation dynamics.

The master equation for the number of $A$'s is
\begin{align}
\frac{\mathrm{d}}{\mathrm{d}t} P_i =& k [- N P_i + (i+1) P_{i+1} + (N-i+1)P_{i-1}] \nonumber\\
&+ s[-2i (N-i)P_i + (i+1)(N-i-1)P_{i+1} \nonumber \\
&\qquad + (i-1) (N-i+1) P_{i-1}],
\end{align}
where $P_i$ is the probability with which the number of $A$'s is $i$.
We can derive the corresponding Fokker-Planck equation:
\begin{align}
\frac{\partial}{\partial t} P =& -k \frac{\mathrm{d}}{\mathrm{d}x}[(N-2x)P]  \nonumber\\
&+ \frac{\mathrm{d}^2}{\mathrm{d}x^2}
\left[ \left( 
k \frac{N}{2} + s x (N-x)
\right) P \right],
\end{align}
and the time-independent solution is obtained as
\begin{align}
P(x) \propto \left[
kN + 2 s x (N-x)
\right]^{k/s-1}
\label{solution_Fokker_Planck}.
\end{align}
The same transition phenomenon discussed in the previous sections
is easily seen to be at $s=k$, which has the uniform steady solution $P_i = 1/(N+1)$.
We note that the change of the volume $V$ in the random walk model in \S~3.1
has the same effect of both changing the number of particles and the rate of one of the reactions.

In the analysis of this effective model, we used the Fokker-Planck equation,
which is based on the assumption of the large system size.
Despite this, the same transition phenomenon can be observed,
which indicate that the discreteness property would not be important for the transition phenomenon.
We will discuss transition phenomenon in more detail in the next section.

\subsection{Discussion}

From the random walk model and the effective model,
we have the following remarks.

First, in order to cause the transition phenomenon discussed by Togashi and Kaneko,
the vanishing effect of the fluctuating force near the boundary, where the number of particles becomes zero,
is important rather than the discreteness property.
It may be difficult to image the effect of the vanishing fluctuating force
by only the numerical experiments in \S~2;
one might consider that only the increase of the internal noise with small $V$ is important.
By the discussion from the random walk model,
it became clear that the decrease of the internal noise near the boundary
is also important for the transition phenomenon.

Secondly, from the analysis of the effective model in \S~3.2,
it is easy to see that this effective model can cause the same transition phenomenon.
In addition, the solution of eq.~\eqref{solution_Fokker_Planck} does not include the discreteness,
because we use the Fokker-Planck equation with the assumption of the large system size.
Then, we can conclude that the discreteness would not be important for the transition phenomenon.

In the absence of noise, there is only one stable fixed point. 
Hence, rate equations, which does not include any noise effects, do not evaluate the transition phenomenon. 
However, the internal noise in the random walk model produces ``quasi-absorbing states''\cite{Assaf2007,Doering2007}
at the $n=0$ and $n=(s_1+s_3)V$ points;
when the random walker in near these boundaries,
the strength of the internal noise becomes weak, so that they tends to stay such region for a long time.
While there is a potential which has the minimum at the fixed point,
the random walker escapes the potential minimum due to the increase of the internal noise,
and it is trapped at the quasi-absorbing states.
From the above discussions, we conclude that the transition phenomenon is caused by 
the quasi-absorbing states.

\section{Conclusions}

In the present paper, we studied the transition phenomenon induced by the internal noise.
We proposed a new chemical reaction network, in which 
the effect of the internal noise is invisible in the deterministic rate equation.
By the numerical experiments, it was confirmed that the chemical reaction network causes
the same transition phenomenon discussed by Togashi and Kaneko \cite{Togashi2001}.
In order to study the transition phenomenon analytically,
the random walk model and the effective model were introduced.
From the discussions for these analytically tractable models,
it was clarified that the quasi-absorbing states, which are produced near the boundaries,
play important roles for the transition phenomenon, rather than the discreteness effects.

The transition phenomenon discussed in the present paper is caused by the quasi-absorbing states,
and we therefore consider that the transition scenario is novel.
The decrease of the volume or the increase of the rate constants 
makes the internal noise large, so that the system escapes from the potential minimum.
On the other hand, there are quasi-absorbing states in which the system tends to stay for a long time.
Hence, the bifurcation of peaks in the probability distribution arises.
In addition, due to the quasi-absorbing states,
one peak does not gradually splits into two peaks,
but the bifurcation of peaks occurs discontinuously.
This behavior of the bifurcation is the same one as the Togashi-Kaneko model.
Because the Togashi-Kaneko model has four chemical substances,
there might be different points between our simple model and the Togashi-Kaneko's one.
In order to clarify the rich phenomena of the Togashi-Kaneko model,
it would be more powerful analytical method.
However, we believe that the concept of the quasi-absorbing states
plays a key role in the transition phenomenon, even if there are a lot of chemical substances.
We expect that this new mechanism for the transition phenomenon will be important
for the study of the statistical physics, especially biophysics.




\begin{thebibliography}{99}
\bibitem{note}
While the surface area of the membrane is proportional to $V^{2/3}$,
for simplicity we assume the rate of the diffusion reaction processes is proportional to $V$;
this assumption alters just the values of the parameters $D_i$ and/or $s_i$.
\end{thebibliography}
\end{document}